\documentclass[aps, prl, twocolumn]{revtex4-2}
\usepackage{amsmath}
\usepackage{amssymb}
\usepackage{enumitem}
\usepackage{graphicx}
\usepackage{hyperref}

\allowdisplaybreaks

\begin{document}
\title{An inequality for relativistic local quantum measurements}

\author{Riccardo Falcone}
\affiliation{Department of Physics, University of Sapienza, Piazzale Aldo Moro 5, 00185 Rome, Italy}

\author{Claudio Conti}
\affiliation{Department of Physics, University of Sapienza, Piazzale Aldo Moro 5, 00185 Rome, Italy}

\begin{abstract}
We investigate the trade-off between vacuum insensitivity and 
sensitivity to excitations in finite-size detectors, 
taking measurement locality as a fundamental constraint. We derive an upper bound on the detectability of vacuum excitation, given a small but nonzero probability of false positives in the vacuum state. The result is independent of the specific details of the measurement or the underlying physical mechanisms of the detector and relies only on the assumption of locality. Experimental confirmation or violation of the inequality would provide a test of the axioms of algebraic quantum field theory, offer new insights into the measurement problem in relativistic quantum physics, and establish a fundamental technological limit in local particle detection.
\end{abstract}

\maketitle

\textit{Introduction}---In relativistic quantum field theory, the vacuum is globally defined as the absence of particles. Yet, when examined locally, it exhibits intense activity, with fluctuations and correlations that make it far from empty~\cite{Reeh1961, SUMMERS1985257, 10.1063/1.527734, Redhead1995, 10.1063/1.533253, Reznik2003, Summers_2011}. 
A key consequence 
is that no local measurement can yield a strictly positive outcome in the vacuum~\cite{Redhead1995}: one cannot construct a pair of local projectors that would correspond to a simple yes/no answer to the question ``Is there no excitation from the vacuum in this region?''. 
This clashes with the intuitive view that detectors register only local excitations while remaining inert on the vacuum.



In this work, we explore the hypothesis that real, finite-size detectors deviate from ideal behavior with respect to the vacuum, in the sense that their vacuum response is small but nonzero. We show that suppressing false positives in the vacuum necessarily imposes constraints on the measurement outcomes for nonvacuum states. Physically, this trade-off can be understood as a consequence of vacuum fluctuations: a detector finely tuned to suppress these fluctuations will inevitably suppress signals from genuine excitations.
We thus unveil a fundamental correlation between responsivity (i.e., the efficiency in converting particles into signals) and dark counts (or false positives)
, stemming directly from the requirement of locality in the underlying quantum measurement.

Concretely, we derive an explicit upper bound on the expectation value of a local measurement in an arbitrary excited state, given its expectation value in the vacuum. 
As an application, we consider scalar-field coherent states localized within the causal completion of the detector region. We obtain a closed-form expression for the bound and compare it with the idealized case of detectors extended over the whole spacetime.


The bound we obtain follows directly from the Reeh–Schlieder theorem \cite{Reeh1961, haag1992local}, which is grounded in the axioms of algebraic quantum field theory (AQFT) and the principle of locality in relativistic quantum theory \cite{haag1992local}. Our results therefore offer a quantitative and falsifiable test of these foundations.

To date, AQFT provides the most rigorous formulation of localization, fully incorporating the principles of quantum mechanics, locality, and relativistic causality. Yet ideal measurements (projective onto an eigenbasis) within finite spacetime regions remain problematic: local algebras lack minimal projections and normal pure states \cite{YNGVASON2005135}, and such idealized measurements have been shown to enable superluminal signaling \cite{sorkin1993impossiblemeasurementsquantumfields, Benincasa_2014, PhysRevD.104.025012}. While several alternative schemes exist \cite{Fewster2020, PhysRevD.104.085014, Papageorgiou2024}, we pursue a simpler hypothesis: local measurements are modeled by infinite-rank projections or by local positive operator-valued measures (POVMs). The inequality we derive provides an experimental test of this assumption.


Our results may also help clarify the degree of locality involved in laboratory quantum measurements. In particular, they address the question of whether the observable associated with a given measurement is confined to a localized region at the detector or represents a more distributed process involving the entire experimental setup
. Since the derived bounds depend explicitly on 
the support of the measurement operator, they provide a means to empirically probe the operational scale at which quantum measurements occur.


No detailed detector modeling is assumed. The result is fully general, relying only on the measurement operator being a local 
POVM element and it culminates in a single inequality derived under minimal assumptions. Such inequalities are particularly valued in quantum foundations \cite{PhysicsPhysiqueFizika.1.195, PhysRevLett.23.880, PhysRevLett.54.857, Emary_2014, RevModPhys.94.045007} because they offer a clear and testable criterion: any empirical violation indicates a breakdown in the underlying theoretical model. If the inequality we present is violated, it could indicate either that the AQFT description of local measurements via localized projections and POVMs is incorrect, or that the presumed localization region must be enlarged, potentially to the entire apparatus or laboratory.

\textit{Ideal and quasi-ideal detectors}---
Consider a detector that clicks in response to excitations above the vacuum. We model its measurement by a 
projection-valued measure (PVM), or more generally, a POVM of the form $\{ \hat{E}_\text{click}, \mathbb{I} - \hat{E}_\text{click} \}$, where a click corresponds to a positive outcome associated with the operator $\hat{E}_\text{click}$. We assume the detector operates within a fixed spacetime region $\mathcal{O}_\text{det}$
. In accordance with the principles of AQFT, the measurement operator $\hat{E}_\text{click}$ must be an element of the local algebra associated with this region, i.e., $\hat{E}_\text{click} \in \mathfrak{A}(\mathcal{O}_\text{det})$.

Hereafter, we denote
$P_\text{click} = \langle \psi | \hat{E}_\text{click} | \psi \rangle$ and $P_\text{dark} = \langle \Omega | \hat{E}_\text{click} | \Omega \rangle$
as the probability that the detector registers a click in an arbitrary state $| \psi \rangle$ and in the vacuum state $| \Omega \rangle$, respectively
. In line with the conventional understanding of particle detectors, an idealized detector is expected to satisfy the following two properties:
\begin{enumerate}[label=(\roman*)]
\item \textbf{Vacuum insensitivity}: The detector should not respond when the system is in the vacuum state, meaning the probability of a dark count should be zero: $P_\text{dark} = 0$. \label{Vacuum_insensitivity}

\item \textbf{Sensitivity to some state}: There exists at least one state $| \psi \rangle$ for which the detector registers a click with nonzero probability, i.e., $P_\text{click} > 0$. 
\label{Sensitivity_to_some_state}
\end{enumerate}

We define a detector as \textit{ideal} if it satisfies both properties \ref{Vacuum_insensitivity} and \ref{Sensitivity_to_some_state} described above \footnote{The notion of ideal detectors, as discussed here, should not be confused with that of ideal quantum measurements. In this context, ideal detectors are defined as those that remain inactive in the vacuum state but produce clicks in certain excited states. In contrast, ideal quantum measurements refer to projective measurements onto the eigenbasis of the corresponding observable.}. In practice, realistic detectors with finite spatial extent cannot simultaneously fulfill both conditions. This limitation arises from the locality of the operator $\hat{E}_\text{click}$ and a corollary of the Reeh--Schlieder theorem
, which states that any positive semi-definite local operator with vanishing vacuum expectation value must be identically zero 
(i.e., the vacuum is separating for any local algebra) \cite{Redhead1995}. Therefore, if a detector satisfies property \ref{Vacuum_insensitivity}---that is, it never clicks in the vacuum---then it must be trivial and cannot satisfy property \ref{Sensitivity_to_some_state}, which requires a nonzero response for some state. 

This no-go result relies on the locality assumption for $\hat{E}_\text{click}$. If this locality condition is lifted, then ideal measurements satisfying both properties \ref{Vacuum_insensitivity} and \ref{Sensitivity_to_some_state} can indeed exist. A notable example is the excitation projector  $\hat{P}_\Omega^\perp = \hat{\mathbb{I}} - | \Omega \rangle \langle \Omega |$, which projects onto the subspace orthogonal to the vacuum. By construction, this operator has zero expectation value in the vacuum and unit expectation value in any state orthogonal to it, thereby satisfying both ideality properties. However, $\hat{P}_\Omega^\perp$ is not an element of any local algebra: it cannot represent a measurement performed by a finite-size detector.


We conjecture that the detectors employed in real experiments are not truly ideal, but rather quasi-ideal, in the sense that they only approximately fulfill properties \ref{Vacuum_insensitivity} and \ref{Sensitivity_to_some_state}. Their response to the vacuum is small but not strictly zero, meaning that $0 < P_\text{dark} \ll 1$. The smaller the value of $P_\text{dark}$ is, the closer the detector is to being effectively ideal with respect to property \ref{Vacuum_insensitivity}. However, property \ref{Vacuum_insensitivity} is fundamentally incompatible with property \ref{Sensitivity_to_some_state}. This leads to a natural trade-off: as a detector becomes more vacuum-insensitive (i.e., as $P_\text{dark}$ decreases), its ability to detect excitations ($P_\text{click}$) also diminishes. In other words, a detector that responds less to the vacuum tends to be less responsive to all states. In what follows, we quantify this trade-off by deriving an upper bound on $P_\text{click}$, under the assumption that $P_\text{dark} \neq 0$.

\textit{Upper bound for} $P_\text{click}$---
The Reeh--Schlieder theorem asserts that any state $| \psi \rangle$ can be approximated arbitrarily well by applying a local operator to the vacuum state $| \Omega \rangle$  (i.e., the vacuum is cyclic for any local algebra) \cite{Reeh1961, haag1992local}. That is, for every $| \psi \rangle $ and spacetime region $\mathcal{O}$, there exists a family of operators $\hat{A}_\zeta $ localized in $\mathcal{O}$ such that the approximation error
\begin{equation}\label{cyclicity}
\mathcal{E}_\zeta  = \left\| | \psi \rangle - \hat{A}_\zeta  | \Omega \rangle  \right\|
\end{equation}
tends to zero as $\zeta \to 0$.

We make use of this property of the vacuum by selecting $\mathcal{O} = \mathcal{O}_\text{det}'$, where $\mathcal{O}_\text{det}'$ denotes the causal complement of the detector region $\mathcal{O}_\text{det}$. This choice ensures that $\hat{A}_\zeta $ is an element of the local algebra $\mathfrak{A}(\mathcal{O}_\text{det}')$, and by microcausality \cite{CLIFTON20011}, it follows that $\hat{A}_\zeta $ commutes with $\hat{E}_\text{click}$
.

Since $\hat{E}_\text{click}$ is a positive operator, it admits a well-defined square root  $\hat{E}_\text{click}^{1/2}$ such that
$P_\text{click} = \| \hat{E}_\text{click}^{1/2} | \psi \rangle \|^2$.
By applying the triangle inequality to this norm, we obtain
\begin{equation}
P_\text{click} \leq \left[ \left\| \hat{E}_\text{click}^{1/2} \left( | \psi \rangle - \hat{A}_\zeta  | \Omega \rangle \right) \right\| + \left\| \hat{E}_\text{click}^{1/2} \hat{A}_\zeta  | \Omega \rangle \right\| \right]^2.
\end{equation}
Since $\hat{E}_\text{click}^{1/2} $ and $ \hat{A}_\zeta $ commute, we can further write
\begin{equation}\label{P_1_2}
P_\text{click}  \leq \left[ \left\| \hat{E}_\text{click}^{1/2} \left( | \psi \rangle - \hat{A}_\zeta  | \Omega \rangle \right) \right\| + \left\| \hat{A}_\zeta  \hat{E}_\text{click}^{1/2} | \Omega \rangle \right\| \right]^2.
\end{equation}

We now invoke the definition of the operator norm, $\| \hat{O} \| = \sup_{| \psi \rangle \neq 0} \| \hat{O} | \psi \rangle \| / \| | \psi \rangle \|$, which holds for any bounded operator $\hat{O}$. This gives the following inequalities:
\begin{equation}
\left\| \hat{E}_\text{click}^{1/2} \left( | \psi \rangle - \hat{A}_\zeta  | \Omega \rangle \right) \right\| \leq \left\| \hat{E}_\text{click}^{1/2} \right\| \left\| | \psi \rangle - \hat{A}_\zeta  | \Omega \rangle \right\|
\end{equation}
and
\begin{equation}
 \left\| \hat{A}_\zeta  \hat{E}_\text{click}^{1/2} | \Omega \rangle \right\| \leq  \left\| \hat{A}_\zeta  \right\| \left\| \hat{E}_\text{click}^{1/2} | \Omega \rangle \right\|.
\end{equation}
Since $\hat{E}_\text{click}$ is a POVM element, we have $\| \hat{E}_\text{click}^{1/2} \| \leq 1 $. Additionally, by using the definitions $\mathcal{E}_\zeta  = \| | \psi \rangle - \hat{A}_\zeta  | \Omega \rangle \|$ and $P_\text{dark} = \langle \Omega |\hat{E}_\text{click} | \Omega \rangle = \| \hat{E}_\text{click}^{1/2} | \Omega \rangle \|^2 $, we obtain
\begin{equation}\label{P_det_1_2_E}
\left\| \hat{E}_\text{click}^{1/2} \left( | \psi \rangle - \hat{A}_\zeta  | \Omega \rangle \right) \right\| \leq  \mathcal{E}_\zeta 
\end{equation}
and
\begin{equation}\label{A_P_1_2_A_sigma}
 \left\| \hat{A}_\zeta  \hat{E}_\text{click}^{1/2} | \Omega \rangle \right\| \leq  \left\| \hat{A}_\zeta  \right\| \sqrt{P_\text{dark}}.
\end{equation}
By applying the bounds from Eqs.~\eqref{P_det_1_2_E} and \eqref{A_P_1_2_A_sigma} to Eq.~\eqref{P_1_2}, we obtain the following inequality:
\begin{equation}\label{upper_bound}
P_\text{click}  \leq \left( \mathcal{E}_\zeta  + \left\| \hat{A}_\zeta  \right\| \sqrt{P_\text{dark}} \right)^2.
\end{equation}

To gain further insight into this bound, consider a state $| \psi \rangle$ prepared by applying a unitary operator $\hat{U}$ to the vacuum $| \Omega \rangle$, i.e., $| \psi \rangle = \hat{U} | \Omega \rangle$. If $\hat{U}$ is localized in the causal complement of the detector region, i.e., $\hat{U} \in \mathfrak{A}(\mathcal{O}_\text{det}')$, then, in accordance with relativistic causality, the detector cannot distinguish the state $| \psi \rangle$ from the vacuum, and thus $P_\text{click} = P_\text{dark}$ \footnote{This follows from microcausality, which ensures that $\hat{U}$ and $\hat{E}_\text{click}$ commute \cite{CLIFTON20011}, and from the unitarity of $\hat{U}$, so that $\langle \Omega | \hat{U}^\dagger \hat{E}_\text{click} \hat{U} | \Omega \rangle = \langle \Omega |  \hat{U}^\dagger  \hat{U} \hat{E}_\text{click} | \Omega \rangle = \langle \Omega | \hat{E}_\text{click} | \Omega \rangle $ \cite{RevModPhys.90.045003}.}. In this case, the bound in Eq.~\eqref{upper_bound} becomes uninformative, offering no further constraint on the relationship between $P_\text{click}$ and $ P_\text{dark}$.

To obtain a nontrivial bound, we instead consider the case where $\hat{U}$ is localized in the causal completion of the detector region, i.e., $\hat{U} \in \mathfrak{A}(\mathcal{O}_\text{det}'')$, with $\mathcal{O}_\text{det}'' = (\mathcal{O}_\text{det}')'$ denoting the double causal complement of $\mathcal{O}_\text{det}$.  This setup describes a scenario in which an agent, confined to the causal completion of the detector, locally prepares the excitation $| \psi \rangle$. Even though $\hat{A}_\zeta $ and $\hat{U}$ are localized in spacelike-separated regions, the state $\hat{A}_\zeta  | \Omega \rangle$ can still approximate $\hat{U} | \Omega \rangle$ to arbitrary precision $\mathcal{E}_\zeta $. As Haag pointed out \cite{haag1992local}, this is due to the vacuum exhibiting weak but nonzero correlations across arbitrarily large spacelike separations. Improving the approximation (i.e., smaller $\mathcal{E}_\zeta $) comes at the cost of increasing the operator norm of $\hat{A}_\zeta $ \footnote{This can be shown by contradiction: suppose the family $\hat{A}_\zeta $ converges to a bounded operator $\hat{A}$. Then, $(\hat{U} - \hat{A}) | \Omega \rangle = 0$, and by the separating property of the vacuum \cite{Redhead1995}, it would follow that $\hat{U} = \hat{A}$. This contradicts the fact that the two operators are localized in disjoint regions. Hence, $\hat{A}_\zeta $ must become unbounded as $\mathcal{E}_\zeta  \to 0$.}. As a result, in Eq.\ (\ref{upper_bound}), a small value of $\mathcal{E}_\zeta$ is necessarily compensated by a large norm $\| \hat{A}_\zeta  \|$.

Crucially, while $P_\text{dark}$ is fixed by the physical properties of the detector, the parameter $\zeta$ remains free to choose. This flexibility allows one to optimize the bound in Eq.~\eqref{upper_bound} by minimizing its right-hand side over all positive values of $\zeta$, leading to the sharpest possible inequality:
\begin{equation}\label{upper_bound_min}
P_\text{click}  \leq \min_{\zeta>0} \left( \mathcal{E}_\zeta  + \left\| \hat{A}_\zeta  \right\| \sqrt{P_\text{dark}}  \right)^2.
\end{equation}

This inequality clearly shows that fulfilling the ideality condition \ref{Vacuum_insensitivity} (vacuum insensitivity) inevitably leads to a violation of the ideality condition \ref{Sensitivity_to_some_state} (sensitivity to certain states). If $P_\text{dark}$ is set to zero, meaning the detector is entirely insensitive to the vacuum, the minimum occurs at $\zeta = 0$, where $\mathcal{E}_\zeta  = 0 $. This leads to the strongest possible constraint on $P_\text{click}$, namely $P_\text{click} = 0$, which implies that the detector is also completely insensitive to the state $| \psi \rangle$.

In the nonideal case where $P_\text{dark}$ is different from zero, the term $\mathcal{E}_\zeta $ in the sum $\mathcal{E}_\zeta  + \| \hat{A}_\zeta  \| \sqrt{P_\text{dark}} $ is always compensated by the second term, which diverges as $\zeta \to 0$.  As a result, the minimum of the sum is achieved through an optimal trade-off between these terms, leading to a bound that remains strictly greater than zero.


\textit{Scalar-field coherent states}---To illustrate with a concrete example, we consider a Klein-Gordon field $\hat{\phi}$ and a coherent state of the form $| f \rangle = \hat{W}(f) | \Omega \rangle$, where $\hat{W}(f) = \exp [i \hat{\phi}(f) ]$ is the Weyl operator associated with the test function $f$, and $\hat{\phi}(f) $ denotes the field operator smeared with $f$. We assume that $\hat{W}(f)$ is localized within the region $\mathcal{O}_\text{det}''$, thereby excluding the trivial case where the state is prepared unitarily in $\mathcal{O}_\text{det}'$. To ensure this localization, we impose that the support of $f$ is entirely contained within $\mathcal{O}_\text{det}''$, i.e., $\operatorname{supp}(f) \subseteq \mathcal{O}_\text{det}''$.

The Reeh--Schlieder theorem guarantees the existence of a family of operators $\hat{A}_\zeta(f)$, localized in the region $\mathcal{O}'_\text{det}$, such that the coherent state $| f \rangle$ can be approximated by $\hat{A}_\zeta(f) | \Omega \rangle$, with approximation error
\begin{equation}
\mathcal{E}_\zeta (f) = \left\| | f \rangle - \hat{A}_\zeta(f) | \Omega \rangle  \right\|
\end{equation}
vanishing as $\zeta \to 0$. 
In our earlier work \cite{falcone2025reehschliederapproximationcoherentstates}, we presented an explicit construction of $\hat{A}_\zeta(f)$ for coherent states. A brief outline is given here, with the full derivation available in \cite{falcone2025reehschliederapproximationcoherentstates}.

We start by selecting a coordinate system $x = (x^0, x^1, x^2, x^3)$ such that the region $\mathcal{O}_\text{det}$ is entirely contained within the left wedge $\mathcal{W}_\text{L}$, defined by the condition $x^1<-|x^0|$. Let $J$ denote the spacetime reflection about the $x^0$ and $x^1$ directions, $J (x^0, x^1, x^2, x^3) = (-x^0, -x^1, x^2, x^3)$, and let $\Lambda_1(\eta)$ represent a Lorentz boost along the $x^1$-direction, $\Lambda_1(\eta)(x^0, x^1, x^2, x^3) = [\cosh (\eta) x^0 + \sinh (\eta) x^1, \cosh (\eta) x^1 + \sinh (\eta) x^0, x^2, x^3]$. Based on these definitions, we construct the family of operators $\hat{A}_\zeta(f)$ as
\begin{equation}\label{W_prime_zeta}
\hat{A}_\zeta(f) = \int_\mathbb{R} d\eta \, G_\zeta(\eta - i \pi) \, \hat{W}\!\left[f \circ J \circ \Lambda_1(-\eta)\right],
\end{equation}
where $G_\zeta(\eta) = \exp (- \eta^2/ 2 \zeta) / \sqrt{2 \pi \zeta}$ is a Gaussian function with variance $\zeta$. Each operator $\hat{A}_\zeta(f)$ is localized in $\mathcal{O}_\text{det}'$, and its action on the vacuum $| \Omega \rangle$ approximates the target coherent state $| f \rangle$ with approximation error
\begin{widetext}
\begin{equation}\label{epsilon_zeta_F}
\mathcal{E}_\zeta (f) = \sqrt{1 - \int_\mathbb{R} d\eta  [ 2 G_\zeta(\eta) -  G_{2 \zeta}(\eta)]  \exp \left\lbrace W_2 \! \left[ f,  f \circ \Lambda_1(\eta) \right] - W_2 (f,f) \right\rbrace },
\end{equation}
\end{widetext}
where $W_2(f_1,f_2) = \langle \Omega | \hat{\phi}(f_1) \hat{\phi}(f_2) | \Omega \rangle$ denotes the two-point Wightman function smeared with test functions $f_1$ and $f_2$. The right-hand side of Eq.~\eqref{epsilon_zeta_F} converges to zero as $\zeta \to 0$, thereby ensuring that $\hat{A}_\zeta(f) | \Omega \rangle$ approximates $| f \rangle$ to arbitrary precision.

An upper bound on the operator norm of $ \hat{A}_\zeta(f) $ can be obtained by applying the triangle inequality to Eq.~\eqref{W_prime_zeta}:
\begin{equation}
\left\| \hat{A}_\zeta(f) \right\| \leq \int_\mathbb{R} d\eta \, |G_\zeta(\eta - i \pi)| \left\| \hat{W}\!\left[f \circ J \circ \Lambda_1(-\eta)\right] \right\| .
\end{equation}
Since every unitary operator has norm $1$, it follows that $\| \hat{W} [f \circ J \circ \Lambda_1(-\eta) ] \| = 1 $, giving
\begin{equation}\label{W_prime_zeta_norm_upper_bound_Gaussian}
\left\| \hat{A}_\zeta(f) \right\| \leq \exp\! \left( \frac{\pi^2}{2\zeta}\right).
\end{equation}
This bound diverges as $\zeta \to 0$, which is consistent with expectations: achieving an increasingly accurate approximation of the state $| f \rangle$ via operators localized in the causal complement $\mathcal{O}_\text{det}'$ requires $\hat{A}_\zeta(f)$ to become unbounded in the limit $\zeta \to 0$.

We denote the probability that the detector clicks in the coherent state $| f \rangle$ as
$P_\text{click}(f) = \langle f | \hat{E}_\text{click} | f \rangle$.
By plugging Eqs.~\eqref{epsilon_zeta_F} and \eqref{W_prime_zeta_norm_upper_bound_Gaussian} into the general bound \eqref{upper_bound}, with the natural identifications $P_\text{click}   \mapsto P_\text{click} (f)$, $\mathcal{E}_\zeta  \mapsto \mathcal{E}_\zeta (f) $ and $\hat{A}_\zeta  \mapsto \hat{A}_\zeta(f)$, we obtain a concrete expression of the inequality for the specific case where 
$| \psi \rangle = | f \rangle$. The resulting bound reads:
\begin{equation}\label{upper_bound_f_O_d_1}
P_\text{click} (f)  \leq \left[ \mathcal{E}_\zeta (f) + \exp\! \left( \frac{\pi^2}{2\zeta}\right) \sqrt{P_\text{dark}} \right]^2.
\end{equation}

This inequality is valid for any positive value of $\zeta$. To obtain the tightest possible bound, we minimize the right-hand side of Eq.~\eqref{upper_bound_f_O_d_1} with respect to $\zeta>0$
:
\begin{equation}\label{upper_bound_f_O_d_2}
P_\text{click} (f)  \leq \min_{\zeta>0} \left[ \mathcal{E}_\zeta (f) + \exp\! \left( \frac{\pi^2}{2\zeta}\right) \sqrt{P_\text{dark}} \right]^2.
\end{equation}


The coordinate system $x$ can still be chosen freely, provided $\mathcal{O}_\text{det}$ remains entirely within $\mathcal{W}_\text{L}$. However, to sharpen the bound in Eq.~\eqref{upper_bound_f_O_d_2}, it is advantageous to select coordinates that optimize the estimate. A general guideline is to choose the coordinates such that $\mathcal{O}_\text{det}$ lies as close as possible to the origin. This choice minimizes the effect of the boost $\Lambda_1(\eta)$ in Eq.~\eqref{epsilon_zeta_F}, thereby reducing the value of $\mathcal{E}_\zeta (f)$. Physically, this corresponds to the fact that the closer is the support of $\hat{A}_\zeta(f) $ to $\mathcal{O}_\text{det}$, the more easily $\hat{A}_\zeta(f) $ can approximate $| f \rangle$.

\textit{A simple model}---As an illustrative toy model, we consider a detector with spherical geometry. Its associated spacetime region, $\mathcal{O}_\text{det}$, is taken to be a four-dimensional hypersphere of radius $R_\text{det}$. We assume that the coherent-state smearing function $f$ is supported in a concentric hypersphere of radius $R_\text{coh} \leq R_\text{det}$. The coordinate system is chosen such that the common center of these hyperspheres is located at the spacetime point $C = (0,-\sqrt{2}R_\text{det},0,0)$, ensuring that $\mathcal{O}_\text{det}$ lies entirely within the left wedge $\mathcal{W}_\text{L}$. Also, we take the field $\hat{\phi}$ to be massless, thereby focusing on the strictly relativistic regime.

We model the coherent state using the smearing function
\begin{equation}\label{f}
f(x) = \frac{\alpha}{R_\text{coh}^3} \, \vartheta\!\left( 1-\frac{\|x - C\|}{R_\text{coh}} \right),
\end{equation}
where $\alpha$ is a real, dimensionless amplitude, $\| \cdot \|$ is the Euclidean norm, and
\begin{align}
& \vartheta(s) =  \frac{\Phi(s)}{\Phi(s) + \Phi(1 - s)}, & \Phi(s) = \begin{cases}
e^{-1/s}, & \text{ if } s>0, \\
0 , & \text{ if } s \leq 0.
\end{cases}
\end{align}
are auxiliary functions that ensures $f$ is smooth and supported within a hypersphere of radius $R_\text{coh}$ centered at $C$.

\begin{figure}[h]
\includegraphics[width=\columnwidth]{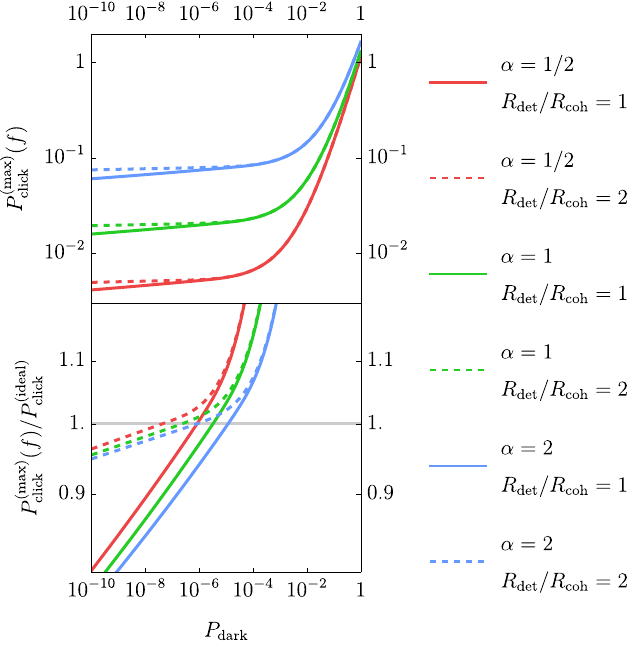}
\caption{Upper panel: upper bound $P_\text{click}^\text{(max)}(f)$ on the click probability $P_\text{click}(f)$ as a function of the dark count $P_\text{dark}$. Lower panel: ratio between the upper bound $P_\text{click}^\text{(max)}(f)$ for a finite-size detector and the click probability $P_\text{click}^\text{(ideal)}$ for an idealized infinitely extended detector. Results are shown for different coherent states $| f \rangle$ and detector geometries. The smearing function $f$ is given by Eq.~\eqref{f}, where $\alpha$ denotes the amplitude and $R_\text{coh}$ the radius of the sphere supporting the coherent state. The detector is modeled as a concentric sphere of radius $R_\text{det}$. The plots are displayed on a log--log scale.}\label{Pmax}
\end{figure}

We numerically evaluate the upper bound  $P_\text{click}^\text{(max)}(f)$, defined as the right-hand side of Eq.~\eqref{upper_bound_f_O_d_2}, for different choices of $\alpha$, $R_\text{det}/R_\text{coh}$, $P_\text{dark}$. Specifically, we consider $\alpha = 1/2, 1, 2$ and $R_\text{det}/R_\text{coh} = 1, 2$, while $P_\text{dark}$ ranges from $10^{-10}$ and $1$. Further details are provided in the Supplemental Material \cite{Supplemental}. The results, shown in the upper panel of Fig.~\ref{Pmax}, reveal that for all parameter choices $\alpha$ and $R_\text{det}/R_\text{coh}$, the bound $P_\text{click}^\text{(max)}(f)$ decreases as $P_\text{dark}$ is reduced. This behaviour reflects the non-ideality of the detector and the intrinsic trade-off between vacuum insensitivity and sensitivity to the coherent state $| f \rangle$: suppressing false positives in the vacuum (i.e., lowering $P_\text{dark}$) inevitably tightens the response of the detector to excitations, thereby reducing the maximum possible value of $P_\text{click}(f)$.

\textit{Comparison with nonlocal measurements}---We emphasize that Eq.~\eqref{upper_bound_f_O_d_2} remains quite general,  as it rests solely on the assumption that the measurement is local. To better appreciate its implications, it is useful to compare this result with a scenario involving a nonlocal measurement. A notable example is the excitation projector $\hat{P}_\Omega^\perp = \hat{\mathbb{I}} - | \Omega \rangle \langle \Omega |$, which perfectly distinguishes the vacuum from any excited state: it gives zero when applied to the vacuum and one for any state orthogonal to it. While this operator appears ideal in its ability to detect excitations, it is fundamentally nonlocal, as it does not belong to any local operator algebra. Consequently, it cannot represent a measurement performed by a finite-size detector.

To make this comparison concrete, let us define the probabilities that an idealized, infinite-size detector clicks in the coherent state $| f \rangle$ and in the vacuum $| \Omega \rangle$ as
$P_\text{click}^\text{(ideal)}(f) = \langle f | \hat{P}_\Omega^\perp | f \rangle$ and $P_\text{dark}^\text{(ideal)} = \langle \Omega | \hat{P}_\Omega^\perp | \Omega \rangle$, respectively.
By the standard inner product for coherent states, $\langle f_1 | f_2 \rangle = \exp [W_2(f_1, f_2) - W_2(f_1, f_1)/2  - W_2(f_2, f_2)/2]$, we find that
\begin{equation}\label{P_Omega_perp_f}
P_\text{click}^\text{(ideal)}(f) = 1 - \exp \! \left[ -W_2 (f,f) \right].
\end{equation}

The ratio $P_\text{click}(f) / P_\text{click}^\text{(ideal)}$ quantifies the deviation of the local measurement from the nonlocal ideal one. An upper bound for this ratio is given by $P_\text{click}^\text{(max)}(f) / P_\text{click}^\text{(ideal)}$. Whenever this bound falls below unity, it follows that $P_\text{click}(f)$ must also lie strictly below $P_\text{click}^\text{(ideal)}$, signaling a definite departure from the ideal case.

As a practical application, we revisit the toy model of a spherically localized coherent state [Eq.~\eqref{f}] with amplitude $\alpha$ and radius $R_\text{coh}$, together with a concentric spherical detector of radius $R_\text{det}$ and dark-count probability $P_\text{dark}$. Using Eq.~\eqref{P_Omega_perp_f}, we compute the ratio between the upper bound $P_\text{click}^\text{(max)}(f)$ on the click probability $P_\text{click}$ for this finite-size setup and the corresponding click probability $P_\text{click}^\text{(ideal)}$ in the idealized limit of an infinitely extended detector. The results, shown in the lower panel of Fig.~\ref{Pmax}, reveal that as the detector size $R_\text{det}$ decreases relative to $R_\text{coh}$, the ratio $P_\text{click}^\text{(max)}(f) / P_\text{click}^\text{(ideal)}$ diminishes, indicating a departure from ideal detector behavior.

\textit{Concluding remarks}---
The main result of this work is the inequality in Eq.~\eqref{upper_bound_min}, which establishes an upper bound on the click probability of a finite-size detector in terms of its dark count rate. This bound depends on the spacetime region $\mathcal{O}_\text{det}$ in which the measurement operator $\hat{E}_\text{click}$ is localized. 

One might raise the objection that actual measurements in laboratory settings are not strictly confined to the detector itself. Rather, the physical process underlying the measurement may extend over a larger region, potentially including parts of the surrounding apparatus. 
This ambiguity has significant implications for 
the inequality in Eq.~\eqref{upper_bound_min}, since the bound for $P_\text{click}$ depends on 
$\mathcal{O}_\text{det}$. If the effective region of the measurement is larger than the physical detector, the corresponding bound could differ significantly.

Interestingly, this opens up the possibility of using Eq.~\eqref{upper_bound_min} as a diagnostic tool: by comparing experimental data against theoretical bounds derived for different choices of $\mathcal{O}_\text{det}$, one could, in principle, infer how localized the measurement truly is. Such comparisons may help shed light on the operational scale at which quantum measurements occur.

These conclusions, of course, rest on the premise that local measurements can be faithfully modeled by local POVMs of the form $\{ \hat{E}_\text{click}, \hat{\mathbb{I}} - \hat{E}_\text{click} \}$. If the region $\mathcal{O}_\text{det}$ accurately captures the operational extent of the measurement, then any empirical violation of the bound would imply that localized POVMs within the AQFT framework fall short of describing the full measurement process. Such a result would offer new insight into the measurement problem in relativistic quantum physics.

\textit{Acknowledgment}---We acknowledge support from HORIZON EIC-2022-PATHFINDERCHALLENGES-01 HEISINGBERG Project No.~101114978.

\bibliography{bibliography}

\begin{thebibliography}{27}%
\makeatletter
\providecommand \@ifxundefined [1]{%
 \@ifx{#1\undefined}
}%
\providecommand \@ifnum [1]{%
 \ifnum #1\expandafter \@firstoftwo
 \else \expandafter \@secondoftwo
 \fi
}%
\providecommand \@ifx [1]{%
 \ifx #1\expandafter \@firstoftwo
 \else \expandafter \@secondoftwo
 \fi
}%
\providecommand \natexlab [1]{#1}%
\providecommand \enquote  [1]{``#1''}%
\providecommand \bibnamefont  [1]{#1}%
\providecommand \bibfnamefont [1]{#1}%
\providecommand \citenamefont [1]{#1}%
\providecommand \href@noop [0]{\@secondoftwo}%
\providecommand \href [0]{\begingroup \@sanitize@url \@href}%
\providecommand \@href[1]{\@@startlink{#1}\@@href}%
\providecommand \@@href[1]{\endgroup#1\@@endlink}%
\providecommand \@sanitize@url [0]{\catcode `\\12\catcode `\$12\catcode
  `\&12\catcode `\#12\catcode `\^12\catcode `\_12\catcode `\%12\relax}%
\providecommand \@@startlink[1]{}%
\providecommand \@@endlink[0]{}%
\providecommand \url  [0]{\begingroup\@sanitize@url \@url }%
\providecommand \@url [1]{\endgroup\@href {#1}{\urlprefix }}%
\providecommand \urlprefix  [0]{URL }%
\providecommand \Eprint [0]{\href }%
\providecommand \doibase [0]{https://doi.org/}%
\providecommand \selectlanguage [0]{\@gobble}%
\providecommand \bibinfo  [0]{\@secondoftwo}%
\providecommand \bibfield  [0]{\@secondoftwo}%
\providecommand \translation [1]{[#1]}%
\providecommand \BibitemOpen [0]{}%
\providecommand \bibitemStop [0]{}%
\providecommand \bibitemNoStop [0]{.\EOS\space}%
\providecommand \EOS [0]{\spacefactor3000\relax}%
\providecommand \BibitemShut  [1]{\csname bibitem#1\endcsname}%
\let\auto@bib@innerbib\@empty
\bibitem [{\citenamefont {Reeh}\ and\ \citenamefont
  {Schlieder}(1961)}]{Reeh1961}%
  \BibitemOpen
  \bibfield  {author} {\bibinfo {author} {\bibfnamefont {H.}~\bibnamefont
  {Reeh}}\ and\ \bibinfo {author} {\bibfnamefont {S.}~\bibnamefont
  {Schlieder}},\ }\bibfield  {title} {\bibinfo {title} {Bemerkungen zur
  unit{\"a}r{\"a}quivalenz von lorentzinvarianten feldern},\ }\href
  {https://doi.org/10.1007/BF02787889} {\bibfield  {journal} {\bibinfo
  {journal} {Il Nuovo Cimento 1955-1965}\ }\textbf {\bibinfo {volume} {22}},\
  \bibinfo {pages} {1051} (\bibinfo {year} {1961})}\BibitemShut {NoStop}%
\bibitem [{\citenamefont {Summers}\ and\ \citenamefont
  {Werner}(1985)}]{SUMMERS1985257}%
  \BibitemOpen
  \bibfield  {author} {\bibinfo {author} {\bibfnamefont {S.~J.}\ \bibnamefont
  {Summers}}\ and\ \bibinfo {author} {\bibfnamefont {R.}~\bibnamefont
  {Werner}},\ }\bibfield  {title} {\bibinfo {title} {The vacuum violates
  {{Bell}}'s inequalities},\ }\href
  {https://doi.org/10.1016/0375-9601(85)90093-3} {\bibfield  {journal}
  {\bibinfo  {journal} {Phys. Lett. A}\ }\textbf {\bibinfo {volume} {110}},\
  \bibinfo {pages} {257} (\bibinfo {year} {1985})}\BibitemShut {NoStop}%
\bibitem [{\citenamefont {Summers}\ and\ \citenamefont
  {Werner}(1987)}]{10.1063/1.527734}%
  \BibitemOpen
  \bibfield  {author} {\bibinfo {author} {\bibfnamefont {S.~J.}\ \bibnamefont
  {Summers}}\ and\ \bibinfo {author} {\bibfnamefont {R.}~\bibnamefont
  {Werner}},\ }\bibfield  {title} {\bibinfo {title} {Bell's inequalities and
  quantum field theory. {{II}}. {{Bell}}'s inequalities are maximally violated
  in the vacuum},\ }\href {https://doi.org/10.1063/1.527734} {\bibfield
  {journal} {\bibinfo  {journal} {J. Math. Phys.}\ }\textbf {\bibinfo {volume}
  {28}},\ \bibinfo {pages} {2448} (\bibinfo {year} {1987})}\BibitemShut
  {NoStop}%
\bibitem [{\citenamefont {Redhead}(1995)}]{Redhead1995}%
  \BibitemOpen
  \bibfield  {author} {\bibinfo {author} {\bibfnamefont {M.}~\bibnamefont
  {Redhead}},\ }\bibfield  {title} {\bibinfo {title} {More ado about nothing},\
  }\href {https://doi.org/10.1007/BF02054660} {\bibfield  {journal} {\bibinfo
  {journal} {Found. Phys.}\ }\textbf {\bibinfo {volume} {25}},\ \bibinfo
  {pages} {123} (\bibinfo {year} {1995})}\BibitemShut {NoStop}%
\bibitem [{\citenamefont {Halvorson}\ and\ \citenamefont
  {Clifton}(2000)}]{10.1063/1.533253}%
  \BibitemOpen
  \bibfield  {author} {\bibinfo {author} {\bibfnamefont {H.}~\bibnamefont
  {Halvorson}}\ and\ \bibinfo {author} {\bibfnamefont {R.}~\bibnamefont
  {Clifton}},\ }\bibfield  {title} {\bibinfo {title} {Generic {{Bell}}
  correlation between arbitrary local algebras in quantum field theory},\
  }\href {https://doi.org/10.1063/1.533253} {\bibfield  {journal} {\bibinfo
  {journal} {J. Math. Phys.}\ }\textbf {\bibinfo {volume} {41}},\ \bibinfo
  {pages} {1711} (\bibinfo {year} {2000})}\BibitemShut {NoStop}%
\bibitem [{\citenamefont {Reznik}(2003)}]{Reznik2003}%
  \BibitemOpen
  \bibfield  {author} {\bibinfo {author} {\bibfnamefont {B.}~\bibnamefont
  {Reznik}},\ }\bibfield  {title} {\bibinfo {title} {Entanglement from the
  vacuum},\ }\href {https://doi.org/10.1023/A:1022875910744} {\bibfield
  {journal} {\bibinfo  {journal} {Found. Phys.}\ }\textbf {\bibinfo {volume}
  {33}},\ \bibinfo {pages} {167} (\bibinfo {year} {2003})}\BibitemShut
  {NoStop}%
\bibitem [{\citenamefont {Summers}(2011)}]{Summers_2011}%
  \BibitemOpen
  \bibfield  {author} {\bibinfo {author} {\bibfnamefont {S.~J.}\ \bibnamefont
  {Summers}},\ }\bibfield  {title} {\bibinfo {title} {Yet more ado about
  nothing: {{The}} remarkable relativistic vacuum state},\ }in\ \href@noop {}
  {\emph {\bibinfo {booktitle} {Deep beauty: {{Understanding}} the quantum
  world through mathematical innovation}}},\ \bibinfo {editor} {edited by\
  \bibinfo {editor} {\bibfnamefont {H.}~\bibnamefont {Halvorson}}}\ (\bibinfo
  {publisher} {Cambridge University Press},\ \bibinfo {address} {Cambridge},\
  \bibinfo {year} {2011})\ pp.\ \bibinfo {pages} {317--342}\BibitemShut
  {NoStop}%
\bibitem [{\citenamefont {Haag}(1992)}]{haag1992local}%
  \BibitemOpen
  \bibfield  {author} {\bibinfo {author} {\bibfnamefont {R.}~\bibnamefont
  {Haag}},\ }\href@noop {} {\emph {\bibinfo {title} {Local quantum physics:
  {{Fields}}, particles, algebras}}},\ \bibinfo {edition} {2nd}\ ed.,\
  Theoretical and {{Mathematical Physics}}\ (\bibinfo  {publisher} {Springer
  Berlin, Heidelberg},\ \bibinfo {year} {1992})\BibitemShut {NoStop}%
\bibitem [{\citenamefont {Yngvason}(2005)}]{YNGVASON2005135}%
  \BibitemOpen
  \bibfield  {author} {\bibinfo {author} {\bibfnamefont {J.}~\bibnamefont
  {Yngvason}},\ }\bibfield  {title} {\bibinfo {title} {The role of type {{III}}
  factors in quantum field theory},\ }\href
  {https://doi.org/10.1016/S0034-4877(05)80009-6} {\bibfield  {journal}
  {\bibinfo  {journal} {Rep. Math. Phys.}\ }\textbf {\bibinfo {volume} {55}},\
  \bibinfo {pages} {135} (\bibinfo {year} {2005})}\BibitemShut {NoStop}%
\bibitem [{\citenamefont
  {Sorkin}(1993)}]{sorkin1993impossiblemeasurementsquantumfields}%
  \BibitemOpen
  \bibfield  {author} {\bibinfo {author} {\bibfnamefont {R.~D.}\ \bibnamefont
  {Sorkin}},\ }\href@noop {} {\bibinfo {title} {Impossible measurements on
  quantum fields}} (\bibinfo {year} {1993}),\ \Eprint
  {https://arxiv.org/abs/gr-qc/9302018} {arXiv:gr-qc/9302018} \BibitemShut
  {NoStop}%
\bibitem [{\citenamefont {Benincasa}\ \emph {et~al.}(2014)\citenamefont
  {Benincasa}, \citenamefont {Borsten}, \citenamefont {Buck},\ and\
  \citenamefont {Dowker}}]{Benincasa_2014}%
  \BibitemOpen
  \bibfield  {author} {\bibinfo {author} {\bibfnamefont {D.~M.~T.}\
  \bibnamefont {Benincasa}}, \bibinfo {author} {\bibfnamefont {L.}~\bibnamefont
  {Borsten}}, \bibinfo {author} {\bibfnamefont {M.}~\bibnamefont {Buck}},\ and\
  \bibinfo {author} {\bibfnamefont {F.}~\bibnamefont {Dowker}},\ }\bibfield
  {title} {\bibinfo {title} {Quantum information processing and relativistic
  quantum fields},\ }\href {https://doi.org/10.1088/0264-9381/31/7/075007}
  {\bibfield  {journal} {\bibinfo  {journal} {Class. Quantum Gravity}\ }\textbf
  {\bibinfo {volume} {31}},\ \bibinfo {pages} {075007} (\bibinfo {year}
  {2014})}\BibitemShut {NoStop}%
\bibitem [{\citenamefont {Borsten}\ \emph {et~al.}(2021)\citenamefont
  {Borsten}, \citenamefont {Jubb},\ and\ \citenamefont
  {Kells}}]{PhysRevD.104.025012}%
  \BibitemOpen
  \bibfield  {author} {\bibinfo {author} {\bibfnamefont {L.}~\bibnamefont
  {Borsten}}, \bibinfo {author} {\bibfnamefont {I.}~\bibnamefont {Jubb}},\ and\
  \bibinfo {author} {\bibfnamefont {G.}~\bibnamefont {Kells}},\ }\bibfield
  {title} {\bibinfo {title} {Impossible measurements revisited},\ }\href
  {https://doi.org/10.1103/PhysRevD.104.025012} {\bibfield  {journal} {\bibinfo
   {journal} {Phys. Rev. D}\ }\textbf {\bibinfo {volume} {104}},\ \bibinfo
  {pages} {025012} (\bibinfo {year} {2021})}\BibitemShut {NoStop}%
\bibitem [{\citenamefont {Fewster}\ and\ \citenamefont
  {Verch}(2020)}]{Fewster2020}%
  \BibitemOpen
  \bibfield  {author} {\bibinfo {author} {\bibfnamefont {C.~J.}\ \bibnamefont
  {Fewster}}\ and\ \bibinfo {author} {\bibfnamefont {R.}~\bibnamefont
  {Verch}},\ }\bibfield  {title} {\bibinfo {title} {Quantum fields and local
  measurements},\ }\href {https://doi.org/10.1007/s00220-020-03800-6}
  {\bibfield  {journal} {\bibinfo  {journal} {Commun. Math. Phys.}\ }\textbf
  {\bibinfo {volume} {378}},\ \bibinfo {pages} {851} (\bibinfo {year}
  {2020})}\BibitemShut {NoStop}%
\bibitem [{\citenamefont {Grimmer}\ \emph {et~al.}(2021)\citenamefont
  {Grimmer}, \citenamefont {Torres},\ and\ \citenamefont
  {{Mart{\'{\i}}n-Mart{\'{\i}}nez}}}]{PhysRevD.104.085014}%
  \BibitemOpen
  \bibfield  {author} {\bibinfo {author} {\bibfnamefont {D.}~\bibnamefont
  {Grimmer}}, \bibinfo {author} {\bibfnamefont {B.~d. S.~L.}\ \bibnamefont
  {Torres}},\ and\ \bibinfo {author} {\bibfnamefont {E.}~\bibnamefont
  {{Mart{\'{\i}}n-Mart{\'{\i}}nez}}},\ }\bibfield  {title} {\bibinfo {title}
  {Measurements in {{QFT}}: {{Weakly}} coupled local particle detectors and
  entanglement harvesting},\ }\href
  {https://doi.org/10.1103/PhysRevD.104.085014} {\bibfield  {journal} {\bibinfo
   {journal} {Phys. Rev. D}\ }\textbf {\bibinfo {volume} {104}},\ \bibinfo
  {pages} {085014} (\bibinfo {year} {2021})}\BibitemShut {NoStop}%
\bibitem [{\citenamefont {Papageorgiou}\ and\ \citenamefont
  {Fraser}(2024)}]{Papageorgiou2024}%
  \BibitemOpen
  \bibfield  {author} {\bibinfo {author} {\bibfnamefont {M.}~\bibnamefont
  {Papageorgiou}}\ and\ \bibinfo {author} {\bibfnamefont {D.}~\bibnamefont
  {Fraser}},\ }\bibfield  {title} {\bibinfo {title} {Eliminating the
  `impossible': {{Recent}} progress on local measurement theory for quantum
  field theory},\ }\href {https://doi.org/10.1007/s10701-024-00756-8}
  {\bibfield  {journal} {\bibinfo  {journal} {Found. Phys.}\ }\textbf {\bibinfo
  {volume} {54}},\ \bibinfo {pages} {26} (\bibinfo {year} {2024})}\BibitemShut
  {NoStop}%
\bibitem [{\citenamefont {Bell}(1964)}]{PhysicsPhysiqueFizika.1.195}%
  \BibitemOpen
  \bibfield  {author} {\bibinfo {author} {\bibfnamefont {J.~S.}\ \bibnamefont
  {Bell}},\ }\bibfield  {title} {\bibinfo {title} {On the {{Einstein Podolsky
  Rosen}} paradox},\ }\href
  {https://doi.org/10.1103/PhysicsPhysiqueFizika.1.195} {\bibfield  {journal}
  {\bibinfo  {journal} {Phys. Phys. Fiz.}\ }\textbf {\bibinfo {volume} {1}},\
  \bibinfo {pages} {195} (\bibinfo {year} {1964})}\BibitemShut {NoStop}%
\bibitem [{\citenamefont {Clauser}\ \emph {et~al.}(1969)\citenamefont
  {Clauser}, \citenamefont {Horne}, \citenamefont {Shimony},\ and\
  \citenamefont {Holt}}]{PhysRevLett.23.880}%
  \BibitemOpen
  \bibfield  {author} {\bibinfo {author} {\bibfnamefont {J.~F.}\ \bibnamefont
  {Clauser}}, \bibinfo {author} {\bibfnamefont {M.~A.}\ \bibnamefont {Horne}},
  \bibinfo {author} {\bibfnamefont {A.}~\bibnamefont {Shimony}},\ and\ \bibinfo
  {author} {\bibfnamefont {R.~A.}\ \bibnamefont {Holt}},\ }\bibfield  {title}
  {\bibinfo {title} {Proposed experiment to test local hidden-variable
  theories},\ }\href {https://doi.org/10.1103/PhysRevLett.23.880} {\bibfield
  {journal} {\bibinfo  {journal} {Phys. Rev. Lett.}\ }\textbf {\bibinfo
  {volume} {23}},\ \bibinfo {pages} {880} (\bibinfo {year} {1969})}\BibitemShut
  {NoStop}%
\bibitem [{\citenamefont {Leggett}\ and\ \citenamefont
  {Garg}(1985)}]{PhysRevLett.54.857}%
  \BibitemOpen
  \bibfield  {author} {\bibinfo {author} {\bibfnamefont {A.~J.}\ \bibnamefont
  {Leggett}}\ and\ \bibinfo {author} {\bibfnamefont {A.}~\bibnamefont {Garg}},\
  }\bibfield  {title} {\bibinfo {title} {Quantum mechanics versus macroscopic
  realism: {{Is}} the flux there when nobody looks?},\ }\href
  {https://doi.org/10.1103/PhysRevLett.54.857} {\bibfield  {journal} {\bibinfo
  {journal} {Phys. Rev. Lett.}\ }\textbf {\bibinfo {volume} {54}},\ \bibinfo
  {pages} {857} (\bibinfo {year} {1985})}\BibitemShut {NoStop}%
\bibitem [{\citenamefont {Emary}\ \emph {et~al.}(2013)\citenamefont {Emary},
  \citenamefont {Lambert},\ and\ \citenamefont {Nori}}]{Emary_2014}%
  \BibitemOpen
  \bibfield  {author} {\bibinfo {author} {\bibfnamefont {C.}~\bibnamefont
  {Emary}}, \bibinfo {author} {\bibfnamefont {N.}~\bibnamefont {Lambert}},\
  and\ \bibinfo {author} {\bibfnamefont {F.}~\bibnamefont {Nori}},\ }\bibfield
  {title} {\bibinfo {title} {Leggett--{{Garg}} inequalities},\ }\href
  {https://doi.org/10.1088/0034-4885/77/1/016001} {\bibfield  {journal}
  {\bibinfo  {journal} {Rep. Prog. Phys.}\ }\textbf {\bibinfo {volume} {77}},\
  \bibinfo {pages} {016001} (\bibinfo {year} {2013})}\BibitemShut {NoStop}%
\bibitem [{\citenamefont {Budroni}\ \emph {et~al.}(2022)\citenamefont
  {Budroni}, \citenamefont {Cabello}, \citenamefont {G{\"u}hne}, \citenamefont
  {Kleinmann},\ and\ \citenamefont {Larsson}}]{RevModPhys.94.045007}%
  \BibitemOpen
  \bibfield  {author} {\bibinfo {author} {\bibfnamefont {C.}~\bibnamefont
  {Budroni}}, \bibinfo {author} {\bibfnamefont {A.}~\bibnamefont {Cabello}},
  \bibinfo {author} {\bibfnamefont {O.}~\bibnamefont {G{\"u}hne}}, \bibinfo
  {author} {\bibfnamefont {M.}~\bibnamefont {Kleinmann}},\ and\ \bibinfo
  {author} {\bibfnamefont {J.-{\AA}.}\ \bibnamefont {Larsson}},\ }\bibfield
  {title} {\bibinfo {title} {Kochen-{{Specker}} contextuality},\ }\href
  {https://doi.org/10.1103/RevModPhys.94.045007} {\bibfield  {journal}
  {\bibinfo  {journal} {Rev. Mod. Phys.}\ }\textbf {\bibinfo {volume} {94}},\
  \bibinfo {pages} {045007} (\bibinfo {year} {2022})}\BibitemShut {NoStop}%
\bibitem [{Note1()}]{Note1}%
  \BibitemOpen
  \bibinfo {note} {The notion of ideal detectors, as discussed here, should not
  be confused with that of ideal quantum measurements. In this context, ideal
  detectors are defined as those that remain inactive in the vacuum state but
  produce clicks in certain excited states. In contrast, ideal quantum
  measurements refer to projective measurements onto the eigenbasis of the
  corresponding observable.}\BibitemShut {Stop}%
\bibitem [{\citenamefont {Clifton}\ and\ \citenamefont
  {Halvorson}(2001)}]{CLIFTON20011}%
  \BibitemOpen
  \bibfield  {author} {\bibinfo {author} {\bibfnamefont {R.}~\bibnamefont
  {Clifton}}\ and\ \bibinfo {author} {\bibfnamefont {H.}~\bibnamefont
  {Halvorson}},\ }\bibfield  {title} {\bibinfo {title} {Entanglement and open
  systems in algebraic quantum field theory},\ }\href
  {https://doi.org/10.1016/S1355-2198(00)00033-2} {\bibfield  {journal}
  {\bibinfo  {journal} {Stud. Hist. Philos. Sci. Part B Stud. Hist. Philos.
  Mod. Phys.}\ }\textbf {\bibinfo {volume} {32}},\ \bibinfo {pages} {1}
  (\bibinfo {year} {2001})}\BibitemShut {NoStop}%
\bibitem [{Note2()}]{Note2}%
  \BibitemOpen
  \bibinfo {note} {This follows from microcausality, which ensures that
  $\protect \hat {U}$ and $\protect \hat {E}_\protect \text {click}$ commute
  \cite {CLIFTON20011}, and from the unitarity of $\protect \hat {U}$, so that
  $\langle \Omega | \protect \hat {U}^\dagger \protect \hat {E}_\protect \text
  {click} \protect \hat {U} | \Omega \rangle = \langle \Omega | \protect \hat
  {U}^\dagger \protect \hat {U} \protect \hat {E}_\protect \text {click} |
  \Omega \rangle = \langle \Omega | \protect \hat {E}_\protect \text {click} |
  \Omega \rangle $ \cite {RevModPhys.90.045003}.}\BibitemShut {Stop}%
\bibitem [{Note3()}]{Note3}%
  \BibitemOpen
  \bibinfo {note} {This can be shown by contradiction: suppose the family
  $\protect \hat {A}_\zeta $ converges to a bounded operator $\protect \hat
  {A}$. Then, $(\protect \hat {U} - \protect \hat {A}) | \Omega \rangle = 0$,
  and by the separating property of the vacuum \cite {Redhead1995}, it would
  follow that $\protect \hat {U} = \protect \hat {A}$. This contradicts the
  fact that the two operators are localized in disjoint regions. Hence,
  $\protect \hat {A}_\zeta $ must become unbounded as $\protect \mathcal
  {E}_\zeta \to 0$.}\BibitemShut {Stop}%
\bibitem [{\citenamefont {Falcone}\ and\ \citenamefont
  {Conti}(2025)}]{falcone2025reehschliederapproximationcoherentstates}%
  \BibitemOpen
  \bibfield  {author} {\bibinfo {author} {\bibfnamefont {R.}~\bibnamefont
  {Falcone}}\ and\ \bibinfo {author} {\bibfnamefont {C.}~\bibnamefont
  {Conti}},\ }\href {https://arxiv.org/abs/2509.09021} {\bibinfo {title}
  {Reeh-schlieder approximation for coherent states}} (\bibinfo {year}
  {2025}),\ \Eprint {https://arxiv.org/abs/2509.09021} {arXiv:2509.09021
  [math-ph]} \BibitemShut {NoStop}%
\bibitem [{Sup()}]{Supplemental}%
  \BibitemOpen
  \href@noop {} {}\bibinfo {note} {See the Supplemental Material for a detailed
  derivation of the upper bound $P_\text{click}^\text{(max)}(f)$ for the
  configuration specified in Eq.~\eqref{f}.}\BibitemShut {Stop}%
\bibitem [{\citenamefont {Witten}(2018)}]{RevModPhys.90.045003}%
  \BibitemOpen
  \bibfield  {author} {\bibinfo {author} {\bibfnamefont {E.}~\bibnamefont
  {Witten}},\ }\bibfield  {title} {\bibinfo {title} {{{APS Medal}} for
  {{Exceptional Achievement}} in {{Research}}: {{Invited}} article on
  entanglement properties of quantum field theory},\ }\href
  {https://doi.org/10.1103/RevModPhys.90.045003} {\bibfield  {journal}
  {\bibinfo  {journal} {Rev. Mod. Phys.}\ }\textbf {\bibinfo {volume} {90}},\
  \bibinfo {pages} {045003} (\bibinfo {year} {2018})}\BibitemShut {NoStop}%
\end{thebibliography}%


\begin{thebibliography}{1}%
\makeatletter
\providecommand \@ifxundefined [1]{%
 \@ifx{#1\undefined}
}%
\providecommand \@ifnum [1]{%
 \ifnum #1\expandafter \@firstoftwo
 \else \expandafter \@secondoftwo
 \fi
}%
\providecommand \@ifx [1]{%
 \ifx #1\expandafter \@firstoftwo
 \else \expandafter \@secondoftwo
 \fi
}%
\providecommand \natexlab [1]{#1}%
\providecommand \enquote  [1]{``#1''}%
\providecommand \bibnamefont  [1]{#1}%
\providecommand \bibfnamefont [1]{#1}%
\providecommand \citenamefont [1]{#1}%
\providecommand \href@noop [0]{\@secondoftwo}%
\providecommand \href [0]{\begingroup \@sanitize@url \@href}%
\providecommand \@href[1]{\@@startlink{#1}\@@href}%
\providecommand \@@href[1]{\endgroup#1\@@endlink}%
\providecommand \@sanitize@url [0]{\catcode `\\12\catcode `\$12\catcode
  `\&12\catcode `\#12\catcode `\^12\catcode `\_12\catcode `\%12\relax}%
\providecommand \@@startlink[1]{}%
\providecommand \@@endlink[0]{}%
\providecommand \url  [0]{\begingroup\@sanitize@url \@url }%
\providecommand \@url [1]{\endgroup\@href {#1}{\urlprefix }}%
\providecommand \urlprefix  [0]{URL }%
\providecommand \Eprint [0]{\href }%
\providecommand \doibase [0]{https://doi.org/}%
\providecommand \selectlanguage [0]{\@gobble}%
\providecommand \bibinfo  [0]{\@secondoftwo}%
\providecommand \bibfield  [0]{\@secondoftwo}%
\providecommand \translation [1]{[#1]}%
\providecommand \BibitemOpen [0]{}%
\providecommand \bibitemStop [0]{}%
\providecommand \bibitemNoStop [0]{.\EOS\space}%
\providecommand \EOS [0]{\spacefactor3000\relax}%
\providecommand \BibitemShut  [1]{\csname bibitem#1\endcsname}%
\let\auto@bib@innerbib\@empty
\bibitem [{\citenamefont {Grafakos}(2014)}]{Grafakos2014}%
  \BibitemOpen
  \bibfield  {author} {\bibinfo {author} {\bibfnamefont {L.}~\bibnamefont
  {Grafakos}},\ }\href {https://doi.org/10.1007/978-1-4939-1194-3} {\emph
  {\bibinfo {title} {Classical Fourier Analysis}}},\ \bibinfo {edition} {3rd}\
  ed.,\ \bibinfo {series} {Graduate Texts in Mathematics}, Vol.\ \bibinfo
  {volume} {249}\ (\bibinfo  {publisher} {Springer},\ \bibinfo {address} {New
  York, NY},\ \bibinfo {year} {2014})\BibitemShut {NoStop}%
\end{thebibliography}%

\end{document}


\title{Supplemental Material: An inequality for finite-size detector measurements}

\author{Riccardo Falcone}
\affiliation{Department of Physics, University of Sapienza, Piazzale Aldo Moro 5, 00185 Rome, Italy}

\author{Claudio Conti}
\affiliation{Department of Physics, University of Sapienza, Piazzale Aldo Moro 5, 00185 Rome, Italy}

\maketitle

In this supplemental material, we consider the toy-model scenario introduced in the main paper, consisting of a four-dimensional, hyperspherical detector region of radius $R_\text{det}$ centered at $C = (0,-\sqrt{2}R_\text{det},0,0)$, together with a massless scalar-field coherent state characterized by the smearing function
\begin{equation}\label{f}
f(x) = \frac{\alpha}{R_\text{coh}^3} \, \chi_\text{coh}(x- C),
\end{equation}
where $\alpha$ is a real, dimensionless amplitude, and $R_\text{coh}$ is the radius of the coherent state. The bump function $ \chi_\text{coh}(x)$ is defined as
\begin{align}
& \chi_\text{coh}(x) = \vartheta\!\left(1 - \frac{\|x \|}{R_\text{coh}} \right), && \vartheta(s) =  \frac{\Phi(s)}{\Phi(s) + \Phi(1 - s)}, && \Phi(s) = \begin{cases}
\exp \!\left( - \frac{1}{s} \right), & \text{ if } s>0, \\
0 , & \text{ if } s \leq 0,
\end{cases}
\end{align}
with $\|\cdot\|$ denoting the Euclidean norm.

The aim of this supplemental material is to derive a more explicit form of the upper bound
\begin{equation}\label{P_click_max}
P_\text{click}^\text{(max)}(f) = \min_{\zeta>0} \left[ \mathcal{E}_\zeta (f) + \exp\! \left( \frac{\pi^2}{2\zeta}\right) \sqrt{P_\text{dark}} \right]^2
\end{equation}
for the click probability $P_\text{click}(f)$, evaluated for different values of $R_\text{det}$, $R_\text{coh}$, $\alpha$ and $P_\text{dark}$. The Reeh–Schlieder error $\mathcal{E}_\zeta (f)$ appearing in Eq.~\eqref{P_click_max} is given by
\begin{equation}\label{epsilon_zeta_F}
\mathcal{E}_\zeta(f) = \sqrt{1 - \int_\mathbb{R} d\eta  [ 2 G_\zeta(\eta) -  G_{2 \zeta}(\eta)]  \exp \left\lbrace W_2 \! \left[ f,  f \circ \Lambda_1(\eta) \right] - W_2 (f,f) \right\rbrace } .
\end{equation}

We start by expanding the field operator $\hat{\phi}$ in terms of Minkowski modes $u$ and their corresponding annihilation operators $\hat{a}$:
\begin{equation}\label{phi_u_a}
\hat{\phi}(x) = \int_{\mathbb{R}^3} d^3 \mathbf{k} \left[ u(\mathbf{k}, x) \hat{a}(\mathbf{k}) + u^*(\mathbf{k},x) \hat{a}^\dagger(\mathbf{k})  \right].
\end{equation}
For simplicity, we allow ourselves the notational shortcut of using $\hat{\phi}(f) = \int_{\mathbb{R}^4} d^4x f(x) \hat{\phi}(x)$ for the smeared field operator and $\hat{\phi}(x)$ for the pointwise field operator. The free modes $u $ are defined as
\begin{equation}\label{u}
u(\mathbf{k}, x) = \left. \frac{e^{ik^\mu x_\mu}}{\sqrt{(2\pi)^3 2 k^0}} \right|_{k^0 = \|\mathbf{k}\|}.
\end{equation}
Here, $\mathbf{k}$ and $\mathbf{x}$ are the spatial components of $k^\mu$ and $x^\mu$, with $k^\mu = (k^0, \mathbf{k})$ and $x^\mu =  (x^0, \mathbf{x})$. The Minkowski product is given by $k^\mu x_\mu = -k^0x^0 + \mathbf{k} \cdot \mathbf{x}$, consistent with the signature $(-,+,+,+)$. The spatial product $\mathbf{k} \cdot \mathbf{x}$ is the standard Euclidean inner product.

The Wightman two-point function smeared with test functions $f_1$ and $f_2$ is defined as
\begin{equation}\label{W_2}
W_2(f_1,f_2) = \left\langle \Omega \middle| \hat{\phi}(f_1) \hat{\phi}(f_2) \middle| \Omega \right\rangle .
\end{equation}
By using Eq.~\eqref{phi_u_a}, together with the canonical commutation relation $[\hat{a}(\mathbf{k}), \hat{a}^\dagger(\mathbf{k}')] = \delta^3(\mathbf{k}-\mathbf{k}')$ and the vacuum condition $\hat{a}(\mathbf{k}) | \Omega \rangle = 0$, we obtain the explicit form
\begin{equation}\label{W_2_f_1_f_2}
W_2(f_1,f_2) = \int_{\mathbb{R}^3} d^3 \mathbf{k} \int_{\mathbb{R}^4} d^4 x \int_{\mathbb{R}^4} d^4 y f_1(x) f_2(y) u(\mathbf{k}, x) u^*(\mathbf{k}, y).
\end{equation}
By setting $f_1 = f$ and $f_2 = f \circ \Lambda_1(\eta)$, this expression becomes
\begin{equation}\label{W_2_f_f_Lambda}
W_2[f,f \circ \Lambda_1(\eta)] = \int_{\mathbb{R}^3} d^3 \mathbf{k} \int_{\mathbb{R}^4} d^4 x \int_{\mathbb{R}^4} d^4 y f(x) f[\Lambda_1(\eta)(y)] u(\mathbf{k}, x) u^*(\mathbf{k}, y).
\end{equation}
Finally, by performing the integration variable transformation $y \mapsto \Lambda_1(-\eta)(y)$, we arrive at
\begin{equation}\label{W_2_f_f_Lambda_2}
W_2[f,f \circ \Lambda_1(\eta)] = \int_{\mathbb{R}^3} d^3 \mathbf{k} \int_{\mathbb{R}^4} d^4 x \int_{\mathbb{R}^4} d^4 y f(x) f(y) u(\mathbf{k}, x) u^*[\mathbf{k}, \Lambda_1(-\eta)(y)].
\end{equation}

By plugging Eqs.~\eqref{f} and \eqref{u} into Eq.~\eqref{W_2_f_f_Lambda_2} and by using the identity $ k^\mu [\Lambda_1(-\eta)(y)]_\mu = [\Lambda_1(\eta)(k)]^\mu y_\mu$, we obtain
\begin{equation}
W_2[f,f \circ \Lambda_1(\eta)] =  \frac{\alpha^2}{(2\pi)^3 2 R_\text{coh}^6} \int_{\mathbb{R}^3} \frac{d^3 \mathbf{k}}{\|\mathbf{k}\|} \int_{\mathbb{R}^4} d^4 x \, \chi_\text{coh}(x- C) \int_{\mathbb{R}^4} d^4 y \, \chi_\text{coh}(y- C) \left. e^{ i k^\mu x_\mu - i [\Lambda_1(\eta)(k)]^\mu y_\mu} \right|_{k^0 = \| \mathbf{k} \|}.
\end{equation}
After shifting the variables $x \mapsto x + C$ and $y \mapsto y + C$, this becomes
\begin{align}\label{W_2_f_f_Lambda_3}
W_2[f,f \circ \Lambda_1(\eta)] = & \frac{\alpha^2}{(2\pi)^3 2 R_\text{coh}^6} \int_{\mathbb{R}^3} \frac{d^3 \mathbf{k}}{\|\mathbf{k}\|} \int_{\mathbb{R}^4} d^4 x \, \chi_\text{coh}(x) \int_{\mathbb{R}^4} d^4 y \, \chi_\text{coh}(y) \nonumber \\
& \times \left. \exp \! \left\lbrace i k^\mu x_\mu - i [\Lambda_1(\eta)(k)]^\mu y_\mu - i \sqrt{2} R_\text{det} k^1 + i \sqrt{2} R_\text{det} [\Lambda_1(\eta)(k)]^1 \right\rbrace \right|_{k^0 = \| \mathbf{k} \|}.
\end{align}

Since $\chi_\text{coh}(x)$ is spherically symmetric, it can be expressed as a radial function $\tilde{\chi}_\text{coh}(\rho)$ defined by
\begin{equation}
\chi_\text{coh}(x) = \tilde{\chi}_\text{coh}\!\left(\frac{\| x \|}{R_\text{coh}} \right).
\end{equation}
The scaling factor $R_\text{coh}$ is introduced so that the argument $\rho$ of $\tilde{\chi}_\text{coh}(\rho)$ is dimensionless. Let
\begin{equation}\label{F_chi}
(\mathcal{F} \chi_\text{coh})(k) = \frac{1}{(2 \pi)^2} \int_{\mathbb{R}^4} d^4x \chi_\text{coh}(x) e^{- i k \cdot x}
\end{equation}
denote the Fourier transform of $\chi_\text{coh}$, where $ k \cdot x $ is the $4$-dimensional Euclidean product. For a spherically symmetric function, the Fourier transform reduces to the one-dimensional integral
\begin{equation}
(\mathcal{F} \chi_\text{coh})(k) = \frac{1}{\| k \|} \int_0^\infty dr \, r^2 \tilde{\chi}_\text{coh}\!\left(\frac{r}{R_\text{coh}} \right) J_1(\| k \| r),
\end{equation}
where $J_1$ is the Bessel function of the first kind of order $1$ (see, e.g., Ref.~\cite{Grafakos2014}, Sec.~B.5 for the general $n$-dimensional formula). By introducing the rescaled function $\tilde{J}_1(s) = J_1(s)/s$ and the dimensionless variable $\rho = r / R_\text{coh}$, this can be rewritten as
\begin{equation}
(\mathcal{F} \chi_\text{coh})(k) = R_\text{coh}^4 \int_0^\infty d\rho \, \rho^3 \tilde{\chi}_\text{coh}( \rho) \tilde{J}_1(R_\text{coh} \| k \| \rho).
\end{equation}
Thus, the Fourier transform takes the form
\begin{equation}\label{F_chi_tilde_F_chi}
(\mathcal{F} \chi_\text{coh})(k) = R_\text{coh}^4 \widetilde{\mathcal{F} \chi}_\text{coh}(R_\text{coh} \| k \|),
\end{equation}
with
\begin{equation}
\widetilde{\mathcal{F} \chi}_\text{coh}( \kappa ) = \int_0^\infty d\rho \, \rho^3 \tilde{\chi}_\text{coh}(\rho) \tilde{J}_1( \kappa  \rho).
\end{equation}

Given the definition in Eq.~\eqref{F_chi} and the radial form in Eq.~\eqref{F_chi_tilde_F_chi}, we have the chain of identities
\begin{equation}
\frac{1}{(2 \pi)^2} \int_{\mathbb{R}^4} d^4x \chi_\text{coh}(x) e^{i k^\mu x_\mu} = (\mathcal{F} \chi_\text{coh})\!\left(k^0,-\mathbf{k}\right) = R_\text{coh}^4 \widetilde{\mathcal{F} \chi}_\text{coh}\!\left[\left\| \left(R_\text{coh} k^0,-R_\text{coh} \mathbf{k}\right) \right\|\right] = R_\text{coh}^4 \widetilde{\mathcal{F} \chi}_\text{coh}(R_\text{coh} \| k \|),
\end{equation}
which allows Eq.~\eqref{W_2_f_f_Lambda_3} to be rewritten as
\begin{align}\label{W_2_f_f_Lambda_5}
W_2[f,f \circ \Lambda_1(\eta)] = & \pi \alpha^2 R_\text{coh}^2 \int_{\mathbb{R}^3} \frac{d^3 \mathbf{k}}{\|\mathbf{k}\|}\widetilde{\mathcal{F} \chi}_\text{coh}\!\left(\sqrt{2} R_\text{coh} \| \mathbf{k} \|\right)    \widetilde{\mathcal{F}\chi}_\text{coh}\!\left[ R_\text{coh} \left\| \Lambda_1(\eta)(\|\mathbf{k}\|, \mathbf{k}) \right\| \right] \nonumber \\
& \times  \left. \exp \! \left\lbrace  - i \sqrt{2} R_\text{det} k^1 + i \sqrt{2} R_\text{det} [\Lambda_1(\eta)(k)]^1 \right\rbrace  \right|_{k^0 = \| \mathbf{k} \|}.
\end{align}

We now switch from Cartesian to spherical coordinates $(k, \theta, \varphi)$ for $\mathbf{k}$, with $k^1 = k \cos(\theta)$, $k^2 = k \sin(\theta) \cos(\varphi)$ and $k^3 = k \sin(\theta) \sin(\varphi)$.  By making the Lorentz boost transformation explicit through $\Lambda_1(\eta)(x^0, x^1, x^2, x^3) = [\cosh (\eta) x^0 + \sinh (\eta) x^1, \cosh (\eta) x^1 + \sinh (\eta) x^0, x^2, x^3]$ and by integrating with respect to $\varphi$, we obtain
\begin{align}\label{W_2_f_f_Lambda_6}
& W_2[f,f \circ \Lambda_1(\eta)] = 2 \pi^2 \alpha^2 R_\text{coh}^2 \int_0^\infty dk \, k \int_0^\pi d\theta \, \sin(\theta) \exp \! \left\lbrace  - i \sqrt{2} R_\text{det} k \left[ \cos (\theta) - \cosh (\eta) \cos (\theta) - \sinh (\eta) \right] \right\rbrace  \nonumber \\
& \times \widetilde{\mathcal{F} \chi}_\text{coh}\!\left(\sqrt{2} R_\text{coh} k\right)  \widetilde{\mathcal{F}\chi}_\text{coh}\!\left\lbrace\sqrt{\left[ \cosh (\eta) + \sinh (\eta) \cos(\theta)\right]^2 + \left[ \cosh (\eta) \cos (\theta) + \sinh (\eta)  \right]^2 + \sin^2 (\theta) } \, R_\text{coh} k\right\rbrace.
\end{align}

To express Eq.~\eqref{W_2_f_f_Lambda_6} in terms of the dimensionless parameters $\alpha$ and $R_\text{det} / R_\text{coh}$, we perform the change of variables $k \mapsto \kappa = k R_\text{coh}$, which gives
\begin{align}\label{W_2_f_f_Lambda_7}
W_2[f,f \circ \Lambda_1(\eta)] = & 2 \pi^2 \alpha^2 \int_0^\infty d\kappa \, \kappa \int_0^\pi d\theta \, \sin(\theta) \exp \! \left\lbrace  - i \sqrt{2} \frac{R_\text{det}}{R_\text{coh}} \kappa \left[ \cos (\theta) - \cosh (\eta) \cos (\theta) - \sinh (\eta) \right] \right\rbrace  \nonumber \\
& \times \widetilde{\mathcal{F} \chi}_\text{coh}\!\left(\sqrt{2}\kappa\right)  \widetilde{\mathcal{F}\chi}_\text{coh}\!\left\lbrace\sqrt{\left[ \cosh (\eta) + \sinh (\eta) \cos(\theta)\right]^2 + \left[ \cosh (\eta) \cos (\theta) + \sinh (\eta)  \right]^2 + \sin^2 (\theta) } \, \kappa \right\rbrace.
\end{align}
In the particular case where $\eta = 0$, this reduces to
\begin{equation}
W_2(f,f) = 2 \pi^2 \alpha^2 \int_0^\infty d\kappa \, \kappa \int_0^\pi d\theta \, \sin(\theta) \left[ \widetilde{\mathcal{F} \chi}_\text{coh}\!\left(\sqrt{2} \kappa\right)\right]^2,
\end{equation}
which simplifies further to
\begin{equation}\label{W_2_f_f}
W_2(f,f) = 4 \pi^2 \alpha^2 \int_0^\infty d \kappa \, \kappa \left[ \widetilde{\mathcal{F} \chi}_\text{coh}\!\left(\sqrt{2} \kappa \right)\right]^2.
\end{equation}

Note that
\begin{equation}
W_2[f,f \circ \Lambda_1(-\eta)] = \{ W_2[f,f \circ \Lambda_1(\eta)]\}^*.
\end{equation}
This follows directly from Eq.~\eqref{W_2_f_f_Lambda_7} by performing the change of variables $\theta \mapsto \pi - \theta$. This observation allows us to simplify Eq.~\eqref{epsilon_zeta_F} 
into
\begin{equation}\label{epsilon_zeta_2}
\mathcal{E}_\zeta (f) = \sqrt{1 - 2 \Re \int_0^\infty d\eta  [ 2 G_\zeta(\eta) -  G_{2 \zeta}(\eta)]  \exp \left\lbrace W_2 \! \left[ f,  f \circ \Lambda_1(\eta) \right] - W_2 (f,f) \right\rbrace }.
\end{equation}

The bound $P_\text{click}^\text{(max)}(f)$ as a function of $\alpha$, $R_\text{det} / R_\text{coh}$ and $P_\text{dark}$ is obtained by plugging Eq.~\eqref{epsilon_zeta_2} into Eq.~\eqref{P_click_max}, together with the boosted Wightman two-point function $W_2[f,f \circ \Lambda_1(\eta)]$ from Eq.~\eqref{W_2_f_f_Lambda_7} and its unboosted counterpart $W_2(f,f)$ from Eq.~\eqref{W_2_f_f}. To generate the plots in Fig.~1 of the main paper, we numerically compute the right hand side of Eqs.~\eqref{W_2_f_f_Lambda_7} and \eqref{W_2_f_f} for various values of $\alpha$ and $R_\text{det} / R_\text{coh}$. We then evaluate $\mathcal{E}_\zeta (f) $ using Eq.~\eqref{epsilon_zeta_2}. Finally, for fixed $P_\text{dark}$ and varying $\zeta$, we find numerically the minimum of $ \mathcal{E}_\zeta (f) + \exp ( \pi^2 / 2 \zeta ) \sqrt{P_\text{dark}}$ which yields the bound $P_\text{click}^\text{(max)}(f)$.

\bibliography{bibliography}